\documentclass{PoS}

\usepackage{amsmath, amsthm, amssymb}
\usepackage{dsfont}
\usepackage{braket}

\newcommand{\Tr}{\mathrm{Tr}}
\def\DLie{\nabla^{a}_{x\mu}}
\def\Dag#1{{#1}^\dagger}
\def\Real#1{\mbox{Re\,}({#1})}

\title{Non perturbative physics from NSPT: renormalons, the gluon condensate and all that}

\ShortTitle{Non perturbative physics from NSPT}

\author{\speaker{Francesco Di Renzo}\\
        Dipartimento di Scienze Matematiche, Fisiche e Informatiche,
        Universit\`a di Parma and INFN,
Gruppo Collegato di Parma, I-43124 Parma, Italy\\
        E-mail: \email{francesco.direnzo@unipr.it}}

\author{Luigi Del Debbio, Gianluca Filaci\\
       Higgs Centre for Theoretical Physics, School of Physics \& Astronomy, University of
Edinburgh, EH9 3FD, U.K\\
       E-mail: \email{luigi.del.debbio@ed.ac.uk, g.filaci@ed.ac.uk}}

\abstract{Numerical Stochastic Perturbation Theory (NSPT) enables very high order computations in Lattice Gauge Theories. We report on the determination of the gluon condensate from lattice QCD measurements of the basic plaquette. This is a long standing problem, which was eventually solved a few years ago in pure gauge. In this context NSPT is crucial: it is actually the only tool enabling the subtraction of the power divergent contribution associated to the identity operator in the OPE for the plaquette. This subtraction
 is actually a delicate issue, since the perturbative expansion of the plaquette is on general ground expected to be an asymptotic one, due to renormalons. This in turn results in ambiguities and the separation of scales in the OPE does not correspond to a separation of perturbative and non-perturbative contributions. All in all, one needs to absorb the
ambiguities attached to the perturbative series into the definition of the condensate itself,
i.e. one needs a prescription. A possible one amounts to summing the perturbative series up to its minimal term, which means computing up to orders which only NSPT can aim at.
Our computation is the first one in QCD, with massless staggered fermions. In order to remove the zero-mode of the gauge field, twisted boundary conditions are adopted for the latter, consistently coupled to fermions in the fundamental representation supplemented with smell degrees of freedom.}

\FullConference{The 36th Annual International Symposium on Lattice Field Theory - LATTICE2018\\
		22-28 July, 2018\\
		Michigan State University, East Lansing, Michigan, USA.}

\begin{document}

\section{An old story: getting the gluon condensate from the OPE of
  the plaquette}

Non-perturbative effects in QCD often arise as power
  corrections. One tool to deal with them was put forward back in the
late seventies by Shifman, Vainshtein and Zakharov and goes under
the name of SVZ sum rules \cite{SVZ}. It amounts basically to an expansion of
correlations functions in the vacuum condensates. One of the latter is
the Gluon Condensate (GC)
\begin{equation}
O_G=-\frac{2}{\beta_0} \frac{\beta(\alpha)}{\alpha}\sum_{a,\mu,\nu}G_{\mu\nu}^a G_{\mu\nu}^a\,,
\end{equation}
which is defined in terms of the beta function
$\beta(\alpha)=\frac{d\alpha}{d\ln\mu}
=-2\alpha \left[\beta_0 (\alpha/4\pi)+\beta_1 (\alpha/4\pi)^2+\dots \right]$ and the field strength
$G_{\mu\nu}^a$. In the following we will often relate the coupling
$\alpha$ to the Wilson action coupling, {\em i.e.}
$\alpha=N_c/(2\pi\beta)$.\\
A non-perturbative determination of $O_G$ has been a 
longstanding challenge for lattice gauge theories. Naively there is a
natural candidate to measure, {\em i.e.} the basic plaquette
\begin{equation}
	\label{eq:PlaqDef}
	P=\frac{1}{6N_cL^4}\sum_{\{P\}}\Real{\Tr\left(1-U_P\right)}\,
\end{equation}
where $U_P$ is the product of the link variables $U_{x\mu}$
around the $1\times 1$ plaquette. 
Indeed in the naive continuum limit
\begin{equation}
	\label{eq:PlaqNaivLim}
a^{-4}P \xrightarrow{a\to0}
\frac{\pi^2}{12N_c}O_G=\frac{\pi^2}{12N_c}\left(
  \frac{\alpha}{\pi}G^2  \right)\,
\;\;\;\;\;\;\;\;
O_G= \frac{\alpha}{\pi}G^2 \left[1+O(\alpha) \right]\,,
\end{equation}
but then mixing with lower dimensional operators takes place (in the
case at hand with the identity operator), which results in a power divergence
\begin{equation}
	\label{eq:plaquetterenormalisation}
	a^{-4}P = a^{-4}Z(\beta)\mathds{1}+\frac{\pi^2}{12N_c}C_G(\beta)O_G+O(a^2\Lambda_\text{QCD}^6)\, .
\end{equation}
If one measures the plaquette by Monte Carlo one in turn has
\begin{equation}
	\label{eq:PlaqMCValue}
	\braket{P}_\text{MC} = Z(\beta) + \frac{\pi^2}{12N_c} C_G(\beta)
	a^4 \braket{O_G} 
	+ O(a^6\Lambda_\text{QCD}^6)\,,
\end{equation}
which can be read as an Operator Product Expansion (OPE): scales are 
separated and Wilson coefficients are computable in
Perturbation Theory (PT)
$$
a^{-1}\gg\Lambda_\text{QCD}\, \quad \quad
	Z(\beta)=\sum_{n=0}p_n\beta^{-(n+1)}\,,
	\quad C_G(\beta)=1+\sum_{n=0}c_n\beta^{-(n+1)}\,,
$$
in which $Z(\beta)$ is sometimes called the {\em perturbative tail}
attached to the plaquette.
(\ref{eq:PlaqMCValue}) has been the starting point for a lattice
determination of the gluon condensate. In principle the recipe is
simple:
\begin{itemize}
\item compute the plaquette by Monte Carlo and in PT;
\item subtract the latter from the former;
\item repeat at different value for the coupling and look for
  asymptotic scaling;
\item read the gluon condensate from the coefficient of the
  contribution scaling as $a^4$.
\end{itemize}
This approach goes back to the work of the Pisa group back in the late
eighties and nineties \cite{PisaGC}. 
\begin{figure}[th]
  \centering
	\includegraphics[height=6cm]{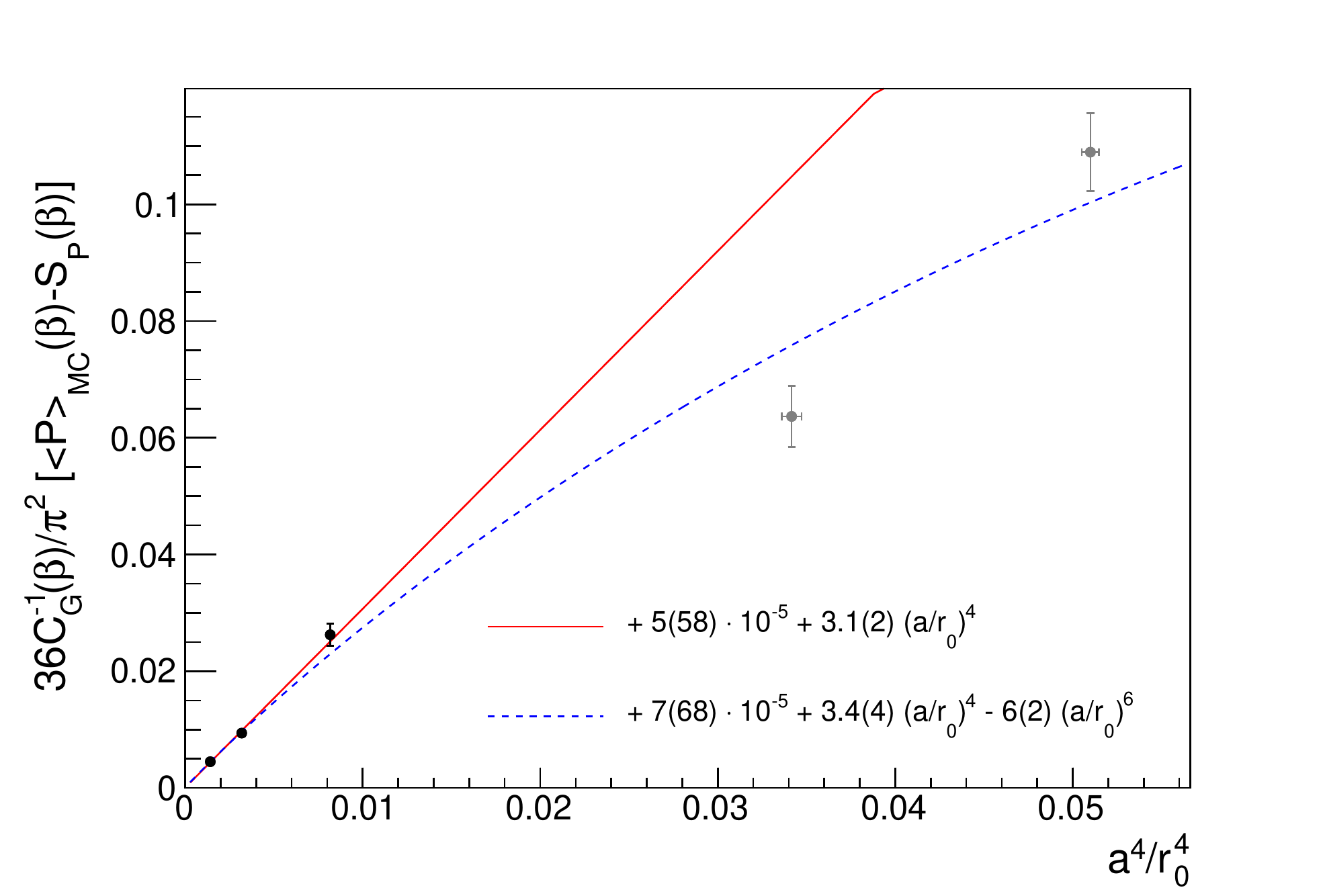}
	\caption{\label{fig:scaling}Scaling of the condensate with
          $a^4$ (solid red line, grey points excluded from the fit), with possibly a $a^6$ contribution
          (dashed blue line, grey points included in the fit). }
\end{figure}
Figure \ref{fig:scaling} shows the result of such a procedure in our
computation, in which we considered $N_f=2$ QCD with (staggered) massless
quarks \cite{DDDRF}. While the procedure appears simple, there are quite a lot of
steps to be taken in order to make sense out of it. All in all, the
subtraction can make sense only in a given prescription and this
ultimately has to do with the fact that the OPE separates the scales,
but not perturbative and non-perturbative contributions. There is
actually a contribution in the perturbative tail attached to the
identity which scales exactly as the gluon condensate. This has to do
with the asymptotic nature of perturbative series in field theories,
which in turn gives rise to ambiguities which in
asymptotically free field theories are known to have to do with the
beta function: this is the famous story of IR renormalons.

\section{IR renormalons}
A dimension 4 condensate is expected to read
$$
W=\int_{0}^{Q^2}\frac{dk^2}{k^2}\left(\frac{k^2}{Q^2}\right)^2 f(k)\,
$$
in which a UV cutoff $Q$ is in place and the form of the integrand is
basically fixed by dimensional and renormalization group
arguments: in particular $f(k)$ can be expressed in terms of the
running coupling. If one picks the simplest expression 
$f(k)=\alpha(k)$ one is left with the expression
\begin{equation}
W^{ren}=\int_{0}^{Q^2}\frac{dk^2}{k^2}\left(\frac{k^2}{Q^2}\right)^2
\alpha(k)\,.
\end{equation}
A quick path into the emergence of (IR) renormalons goes through the
change of variable
\begin{equation}
z=z_0\left(1-\frac{\alpha(Q)}{\alpha(k)}\right) 
\quad z_0 \equiv \frac{(4 \pi)^2}{3\beta_0}
\end{equation}
which leaves us with 
\begin{equation}
\label{eq:Wren_z}
W^{ren}={\cal N} \int_{0}^{\infty} dz \, e^{-\beta z}(z-z_0)^{-1-\gamma}
\end{equation}
where $\gamma \equiv 2 \beta_1/\beta_0^2$.
and a Borel integral can be recognized.  
In the previous expressions the beta function coefficients $\beta_0,\beta_1$
appear; we are retaining only these two terms in the
expansion of the beta function, which is involved in the change of
variable. (\ref{eq:Wren_z}) directly encodes the perturbative
behaviour
\begin{equation}
W^{ren}= \sum_{l=1} \beta^{-l} \left[c_l^{ren} +
  {\cal O}(e^{-z_0\beta}) \right]
\quad c_l^{ren} = {\cal N}' \, \Gamma(l+\gamma) \, z_0^{-l}\, .
\end{equation}
All in all, we can say the following
\begin{itemize}
\item The coefficients of the expansion grow factorially.
\item The integral has in turn a branch cut on the real axis and one
  needs a prescription to compute it. Typically this asks for a detour
  in the complex plane and as a result one picks
  up an imaginary part proportional to $e^{-\beta z_0}$.
\item In order to sum the series one also needs a prescription, with
  an ambiguity which turns out to be just of the same order.
\item The ambiguity we are left with scales just as the GC, {\em i.e.} as
$e^{-\beta z_0} \sim \frac{\Lambda^4}{Q^4}\, .$
\end{itemize}
In view of all this, we need a prescription 
in order to perform the subtraction we are concerned with. But before
getting into this, one question is in order: can we actually detect
the divergent perturbative behaviour we are ready to take care of?

\section{Numerical Stochastic Perturbation Theory}

\begin{figure}[tbp]
	\centering
	\includegraphics[width=.49\textwidth,trim={0.1cm 0 1.4cm 0},clip]{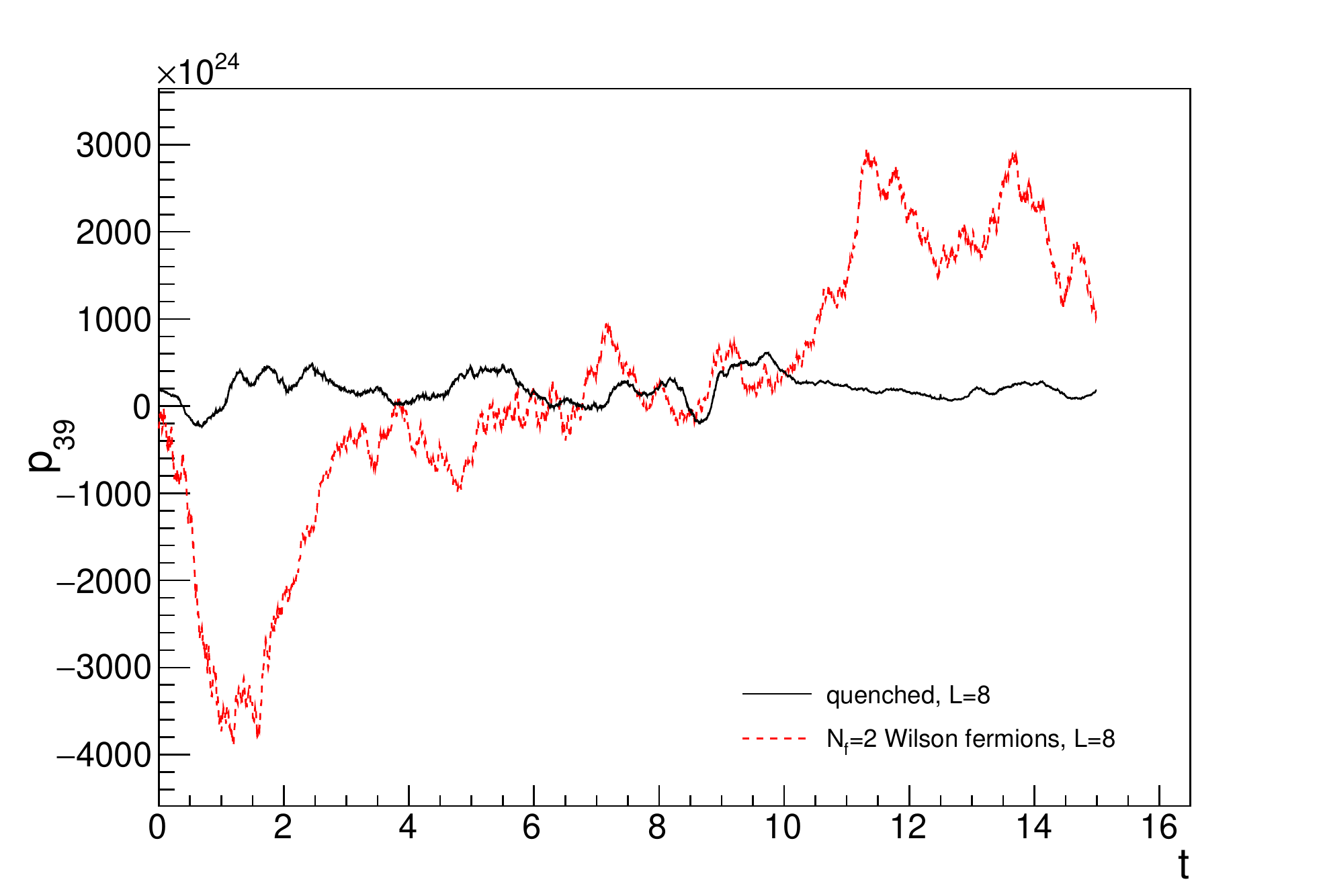}
	\hfill
	\includegraphics[width=.49\textwidth,trim={0.1cm 0 1.4cm 0},clip]{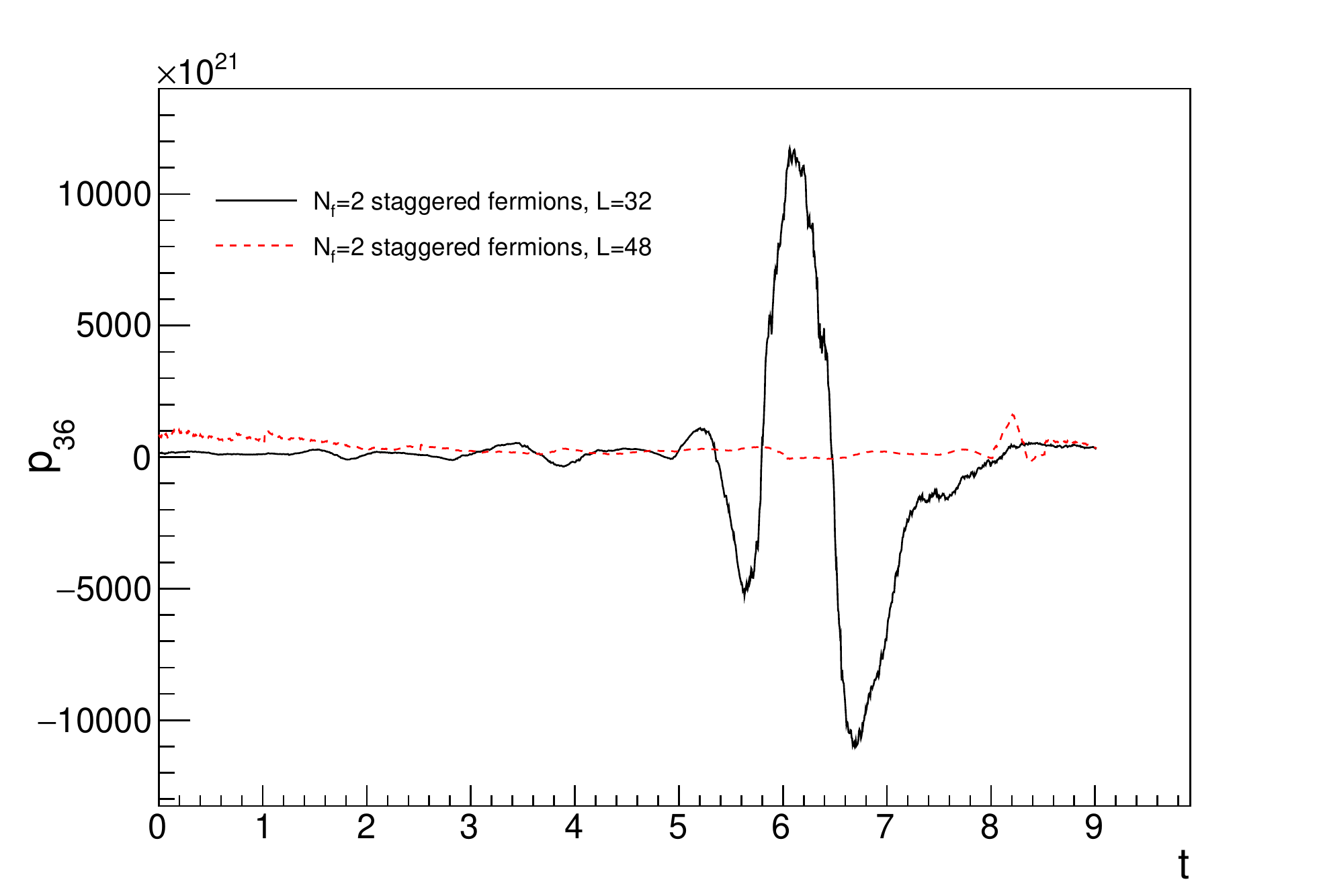}
	\caption{\label{fig:spikes}Examples of spikes at high orders.
          Left ($\beta^{-39}$): quenched vs
          unquenched (Wilson fermions); right ($\beta^{-36}$): $48^4$ vs $32^4$
          (staggered fermions). }
\end{figure}

Explicitly inspecting an asymptotic behaviour asks for the computation of high
orders. Numerical Stochastic Perturbation Theory \cite{NSPT,NSPTferm} (NSPT) is the numerical tool enabling
such a computation. In its original formulation (the one we adhere to)
the method amounts to the numerical implementation of Stochastic
Perturbation Theory as originally introduced in the context of
Stochastic Quantization \cite{ParisiWu}. For a given formulation of a lattice gauge
theory (with gauge group $SU(N_c)$) based on an action $S[U]$, the link
variable $U_{x\mu}$ evolves in the (fictitious) stochastic time $t$
according to the Langevin equation
\begin{equation}
\label{eq:langevin}
\frac{\partial}{\partial t} U_{x\mu}(t;\eta) \, = \, 
\left( -i \nabla_{x\mu} S[U] -i \eta_{x\mu}(t) \right) U_{x\mu}(t;\eta)
\end{equation}
where $\eta_{x\mu}$ is a gaussian noise,
{\em i.e.} $\langle \eta^a(t) {\rangle}_\eta \, = \, 0$
and $\langle \eta_{x\mu}^a(t) \, \eta_{y\nu}^b(t') {\rangle}_\eta \, = \, 
2 \, \delta^{ab} \, \delta_{\nu\mu}\, \delta_{y x} \, \delta(t-t')$.
In our case we have in place both the gauge action $S_G[U]$ and
the determinant of the Dirac operator $M[U]$, resulting in an effective
action $S_{eff}[U]=S_G[U]-\Tr\ln M[U]$. Asymptotically in the
stochastic time, averages over the stochastic noise reconstruct the
path integral averages one is interested in, {\em i.e.}
\begin{equation}
\label{eq:lange2pathint}
\lim_{t\rightarrow\infty} \langle O[U(t;\eta)] {\rangle}_\eta \, =
\frac{1}{Z} \int DU \, e^{- S_{eff}[U]} \, O[U]\, .
\end{equation}
NSPT is obtained by expanding the solution to (\ref{eq:langevin}) in
powers of the coupling
\begin{equation}
\label{eq:NSPTexpansion}
U_{x\mu}(t;\eta) = 1+ \sum_{k=1} \beta^{-k/2}
U^{(k)}_{x\mu}(t;\eta)\, .
\end{equation}
Plugging (\ref{eq:NSPTexpansion}) into (\ref{eq:langevin}) results in a hierarchy of equations,
exactly truncable at any given order. (\ref{eq:lange2pathint}) holds
in turn order-by-order and generates the perturbative expansions
one is aiming at. We notice that explicitly one has
\begin{equation}
\label{eq:nablaS}
\nabla_{x\mu}S = \sum_a T^a \nabla^{a}_{x\mu} S \;\;\;\;\;\;\;\;\;\;\;\;
\DLie S_{eff} = \DLie S_G - \DLie \Tr \ln M  = \DLie S_G - \Tr ( (\DLie M) M^{-1})
\end{equation}
where the second (fermionic) contribution can be obtained via a
stochastic estimator (multi-indices are in place)
\begin{equation}
\label{eq:fermioNoise}
\Real{\Dag{\xi_k}(\DLie M)_{kl}(M^{-1})_{ln}\xi_n}
\end{equation}
in which a second noise has been introduced satisfying
$ \langle \xi_i \xi_j \rangle_{\xi} = \delta_{ij} $.
A numerical integration scheme is needed, {\em e.g.} Euler. This is
not yet the end of the story, since a stochastic gauge fixing is
needed to tame gauge modes \cite{NSPT,NSPTferm}. \\
To avoid zero modes we use twisted boundary conditions (a choice
first made in NSPT in  \cite{GunnarAntonio_0})
\begin{equation}
\label{eq:withStochGaugeFix}
  U_\mu(x+L\hat\nu)=\Omega_\nu
  U_\mu(x)\Omega_\nu^\dag\, 
\quad
  \Omega_\nu\Omega_\mu =
  z_{\mu\nu} \Omega_\mu\Omega_\nu\,\qquad z_{\mu\nu}\in Z_{N_c}
\end{equation}
and consistently give fermions (in the fundamental representation)
smell degrees of freedom, {\em i.e.} copies which transform into each other
according to the anti-fundamental representation of the gauge group
(physical observables are singlets)
\begin{equation}
\label{eq:withStochGaugeFix}
  \psi(x+L\hat\nu)_{ir}=\sum_{j,s}\big(\Omega_\nu\big)_{ij}
  \psi(x)_{js}\big(\Omega_\nu^\dag\big)_{s 
    r}\, . 
\end{equation}
After an exploratory study of critical mass for Wilson fermions, we
eventually chose staggered fermions (for the first time introduced in
NSPT); an obvious choice, given our aiming at high orders.
In a preliminary phase we made use of an updated version
of the PRLGT code \cite{NSPTcode}, while production runs were performed in a new
GRID \cite{grid} NSPT environment\footnote{We thank P. Boyle and G. Cossu for
  their support.}.\\
Very high orders for toy models are known to display numerical
instabilities \cite{NSPTtoy} in NSPT. On the other side these had never been reported
for gauge theories (not even for high orders) in the quenched
case \cite{GunnarAntonio}. With fermions in place, we indeed found instabilities at high
orders: see Figure~\ref{fig:spikes}.
One should keep in mind that in NSPT physical observables do admit an
asymptotic limit; still, we do not know that much on variances (which 
are not connected to physical observables). As expected on general
grounds, the problems are milder on bigger volumes (see Figure
\ref{fig:spikes}). On one side, it is reassuring to notice that 
once a fluctuation takes place, the restoring force eventually takes
the signal back around its average value; on the other side, this can be
not the case if fluctuations are frequent enough to
loose the signal.

\section{Numerical results}

\begin{figure}[bp]
	\centering
	\includegraphics[height=5.6cm]{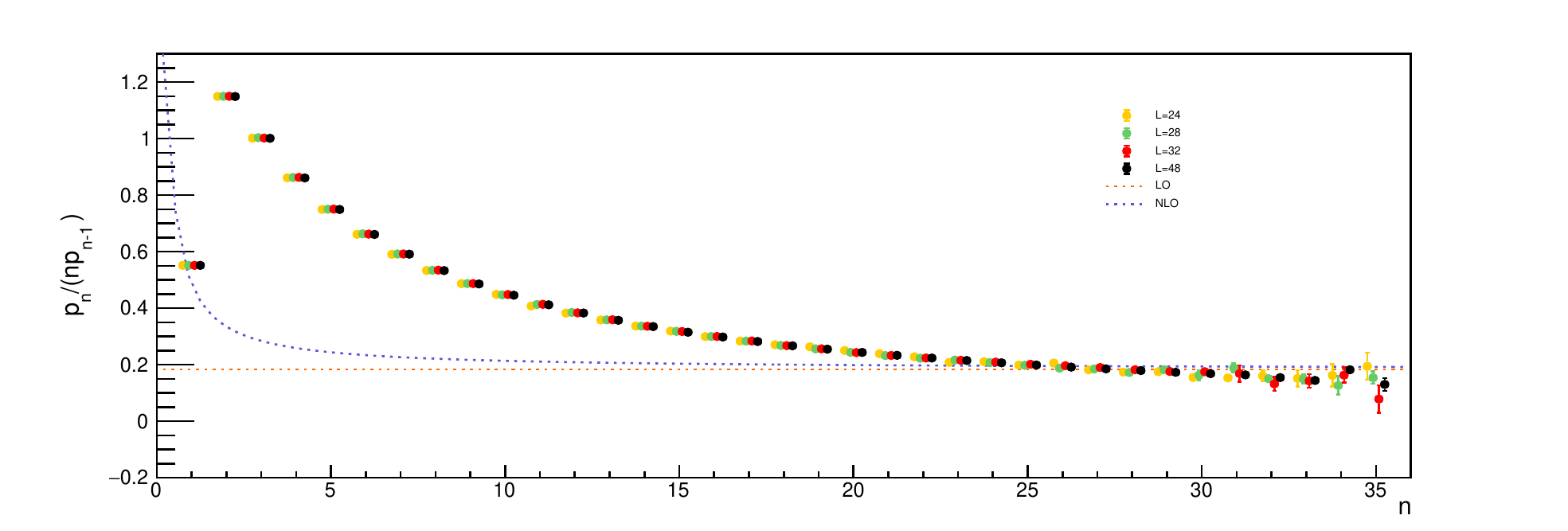}
	\caption{\label{fig:factgrowth}Ratio $p_n/(np_{n-1})$
          extracted from our data at $L=24$, $28$, $32$, $48$ compared
          to leading order (LO) and next-to-leading order (NLO)
          corrections: see Eq (\ref{eq:asymptoticbehaviour}). In order to be visible, points  referring to different volumes are placed side by side.}
\end{figure}

Detecting the renormalon growth has been a challenge for NSPT since the
very early days of the method \cite{8loops}. In recent years, the problem was
successfully solved in the pure gauge case \cite{GunnarAntonio}. One
needs to inspect the ratio
\begin{equation}
	\label{eq:asymptoticbehaviour}
	\frac{p_n}{np_{n-1}} = \frac{3\beta_0}{16\pi^2}
	\left[1+\frac{2\beta_1}{\beta_0^2}\frac{1}{n} + 
	O\left(\frac{1}{n^2}\right)\right]\,,
\quad \mbox{where} \quad
	\braket{P}_\text{pert}=\sum_{n=0}^\infty p_n\,\beta^{-(n+1)}\, .
\end{equation}
This is depicted in Figure~\ref{fig:factgrowth}. Within errors 
the asymptotic behaviour appears to show up, even if it is hard to get
a definite answer: do the points
really flatten to lay on the horizontal line (the signature of renormalons)? 
Our points cross the line, but finite size effects are there.
A definite answer can only come from a careful analysis of
finite size effects: the latter have already been studied in the literature
 \cite{a10,GunnarAntonio}, a study that we do plan to repeat for the case at hand. Having
said this, we do see an inversion point when we sum the
series: although eventually regularised by the finite volume, the
series still appears asymptotic at the orders we are taking into
account. \\
Given the renormalon divergence, one needs a prescription 
to sum the series and a natural one is just summing up to the minimal term (inversion
point), which occurs at a given order ${\bar n}$. The final
recipe to look for the gluon condensate thus amounts to
\begin{equation}
	S(\beta)_P=\sum_{n=0}^{\bar n}p_n\beta^{-(n+1)}\, \;\;\;\;\;\;\;\;\;\;\;\;\;\;\;
	\braket{O_G}=\frac{36}{\pi^2} \,
        C^{-1}_G(\beta)\,a^{-4}\,[\braket{P}_\text{MC}(\beta)-S_P(\beta)]\, .
\end{equation}
Truncating the series is one of the possible prescriptions to sum a
divergent series. An intrinsic ambiguity can be defined as the
imaginary part of the Borel integrand. For the GC this results in
\begin{equation}
	\delta\braket{O_G}=\frac{36}{\pi^2} \, C^{-1}_G(\beta)\,a^{-4}\, \sqrt{\frac{\pi\bar n}{2}} \, p_{\bar n}\beta^{-\bar n -1}
\end{equation}
We stress that any prescription to sum the series amounts to having the
ambiguity absorbed in the definition of the GC in the OPE. One should
always bear this in mind when inspecting results like the one shown in
Figure~\ref{fig:scaling}.

\end{document}